\documentclass[showpacs,preprintnumbers,
amsmath,amssymb,aps,prd]{revtex4}
\usepackage{amsfonts}
\usepackage{amsfonts,graphics,epsfig,subfigure}

\begin{document}

\title{Phase transition and thermodynamic geometry of $f(R)$ AdS black holes in the grand canonical ensemble}
\author{Gu-Qiang Li \footnote{zsgqli@hotmail.com}, Jie-Xiong Mo \footnote{Corresponding author: mojiexiong@gmail.com}}
 \affiliation{Institute of Theoretical Physics, Lingnan Normal University, Zhanjiang, 524048, Guangdong, China}

\begin{abstract}
The phase transition of four-dimensional charged AdS black hole solution in the $R+f(R)$ gravity with constant curvature is investigated in the grand canonical ensemble, where we find novel characteristics quite different from that in canonical ensemble. There exists no critical point for $T-S$ curve while in former research critical point was found for both the $T-S$ curve and $T-r_+$ curve when the electric charge of $f(R)$ black holes is kept fixed. Moreover, we derive the explicit expression for the specific heat, the analog of volume expansion coefficient and isothermal compressibility coefficient when the electric potential of $f(R)$ AdS black hole is fixed. The specific heat $C_\Phi$ encounters a divergence when $0<\Phi<b$ while there is no divergence for the case $\Phi>b$. This finding also differs from the result in the canonical ensemble, where there may be two, one or no divergence points for the specific heat $C_Q$. To examine the phase structure newly found in the grand canonical ensemble, we appeal to the well-known thermodynamic geometry tools and derive the analytic expressions for both the Weinhold scalar curvature and Ruppeiner scalar curvature. It is shown that they diverge exactly where the specific heat $C_\Phi$ diverges.

\end{abstract}

  \maketitle

\section{Introduction}
$f(R)$ gravity has various applications in both gravitation and cosmology. For example, it mimics the cosmological history successfully. One can read the nice reviews~\cite{Felice,Capozziello,Odintsov1} to gain an comprehensive understanding. Believing that black holes in $f(R)$ gravity distinguish from those in Einstein gravity, both the black hole solutions in $f(R)$ gravity and their thermodynamics~\cite{Dombriz}-\cite{xiong6} have received considerable attention.

In our recent paper \cite{xiong6}, we investigated the phase transition of four-dimensional charged AdS black hole solution in the $R+f(R)$ gravity with constant curvature~\cite{Moon98} in the canonical ensemble. To provide a consistent and unified picture of its critical phenomena, we studied not only the critical point of $T-S$ curve and $T-r_+$ curve, but also the divergent behavior of specific heat at constant charge and scalar curvature of Quevedo's geometrothermodynamics \cite{Quevedo2}.

   In this paper, we would like to generalize our recent research \cite{xiong6} to the grand canonical ensemble. This generalization is of interest and it is believed that the phase transition in the grand canonical ensemble will behave quite differently from that in the canonical ensemble, which has been witnessed in our former research of charged topological black holes in Ho\v{r}ava-Liftshitz gravity \cite{jiexiong1} and Lovelock Born-Infeld gravity \cite{jiexiong2} and has also been witnessed in many other references. So the motivation is to probe novel characteristics of phase transition for four-dimensional charged AdS black hole solution in the $R+f(R)$ gravity with constant curvature from the perspective of grand canonical ensemble. Our research will also disclose interesting properties due to $f(R)$ gravity.

    The organization of this paper is as follows. In Sec.~\ref{sec:2} we will review briefly four-dimensional charged AdS black hole solution in the $R+f(R)$ gravity with constant curvature. In Sec.~\ref {sec:3} we will investigate the behavior of temperature and phase transition in grand canonical ensemble. In Sec.~\ref {sec:4}, we will study both the Weinhold thermodynamic geometry \cite{Weinhold} and Ruppeiner thermodynamic geometry \cite{Ruppeiner} in grand canonical ensemble. Conclusions will be drawn in Sec.~\ref {sec:5}.

\section{Review of black hole solution in the $R+f(R)$ gravity with constant curvature}
\label{sec:2}
 In Ref.~\cite{Moon98}, four-dimensional charged AdS black hole solution in the $R+f(R)$ gravity with constant curvature was obtained with its thermodynamic quantities, such as energy, entropy, heat capacity and Helmhotz free energy discussed. $P-V$ criticality of this solution was investigated in Ref.~\cite{Chen}. Recently, we investigated the coexistence curve and the number densities of black hole molecules for this black hole solution~\cite{xiong5} and studied its phase transition in the canonical ensemble when the charge of the black hole is fixed~\cite{xiong6}.

  The corresponding black hole solution reads~\cite{Moon98}
\begin{equation}
ds^2=-N(r)dt^2+\frac{dr^2}{N(r)}+r^2(d\theta^2+sin^2\theta d\phi^2),\label{1}
\end{equation}%
where
\begin{eqnarray}
N(r)&=&1-\frac{2m}{r}+\frac{q^2}{br^2}-\frac{R_0}{12}r^2,\label{2}
\\
b&=&1+f'(R_0).\label{3}
\end{eqnarray}%
In the above solution, $b>0,R_0<0$. Note that the black hole solution reduces to RN-AdS black hole when $b=1,R_0=-12/l^2$.

The black hole ADM mass $M$ and the electric charge $Q$ are related to the parameters $m$ and $q$ respectively as~\cite{Moon98}
\begin{equation}
M=mb,\;\;\; Q=\frac{q}{\sqrt{b}}.\label{4}
\end{equation}%

Its thermodynamic quantities were reviewed in Ref.~\cite{Chen} as follows
\begin{eqnarray}
T&=&\frac{N'(r_+)}{4\pi}=\frac{1}{4\pi r_+}(1-\frac{q^2}{br_+^2}-\frac{R_0r_+^2}{4}).\label{6}
\\
S&=&\pi r_+^2b.\label{7}
\\
\Phi&=&\frac{\sqrt{b}q}{r_+}.\label{8}
\end{eqnarray}
$T$, $S$ and $\Phi$ denote the Hawking temperature, the entropy and the electric potential respectively. Note that the entropy here was derived from the Wald method \cite{Moon98, Felice}. Readers who are interested in it can further read Section 13.2 of reference \cite{Felice} and the famous literature \cite{Wald}.

\section{Phase transition of $f(R)$ AdS black hole in grand-canonical ensemble}
\label{sec:3}
To facilitate the calculation of relevant quantities, it is convenient to reexpress the Hawking temperature as the function of entropy and electric potential

\begin{equation}
T=\frac{4b^2\pi-bR_0 S-4\pi\Phi^2}{16\pi^{3/2}b^{3/2}\sqrt{S}}.\label{9}
\end{equation}
When $b=1,R_0=-12/l^2=4\Lambda$, Eq.(\ref{9}) reduces to
\begin{equation}
T=\frac{\pi-\Lambda S-\pi\Phi^2}{4\pi^{3/2}\sqrt{S}},\label{99}
\end{equation}
which is in accord with the result of RN-AdS black holes \cite{Banerjee1,Banerjee2}.

With Eq.(\ref{9}), it is quite easy to obtain
\begin{eqnarray}
\left(\frac{\partial T}{\partial S}\right)_\Phi&=&\frac{-b(4b\pi+R_0S)+4\pi\Phi^2}{32\pi^{3/2}(bS)^{3/2}}
,\label{10}
\\
\left(\frac{\partial^2 T}{\partial S^2}\right)_\Phi&=&\frac{12b^2\pi+bR_0 S-12\pi\Phi^2}{64\pi^{3/2}b^{3/2}S^{5/2}}
.\label{11}
\end{eqnarray}
The solution for $\left(\frac{\partial T}{\partial S}\right)_\Phi=0$ can be derived as
\begin{equation}
S_1=\frac{-4\pi(b^2-\Phi^2)}{bR_0}.\label{12}
\end{equation}
Note that $b>0, R_0<0$, the condition $0<\Phi<b$ should be satisfied to ensure that the entropy in Eq.(\ref{12}) is positive. For the case $\Phi>b$, no meaningful root satisfies the equation $\left(\frac{\partial T}{\partial S}\right)_\Phi=0$.
Substituting Eq.(\ref{12}) into (\ref{11}), one can obtain that
\begin{equation}
\left(\frac{\partial^2 T}{\partial S^2}\right)_\Phi\mid_{S=S_1}=\frac{bR_0^4}{256\pi^3[R_0(-b^2+\Phi^2)]^{3/2}}>0.\label{13}
\end{equation}
So there is no critical point for $T-S$ curve. This finding differs from our former research, where we found critical point for both the $T-S$ curve and $T-r_+$ curve when the electric charge of $f(R)$ black holes is kept fixed~\cite{xiong6}, providing one more example that the thermodynamics in the grand canonical ensemble is quite different from that in the canonical ensemble.

\begin{figure}
\centerline{\subfigure[]{\label{1a}
\includegraphics[width=8cm,height=6cm]{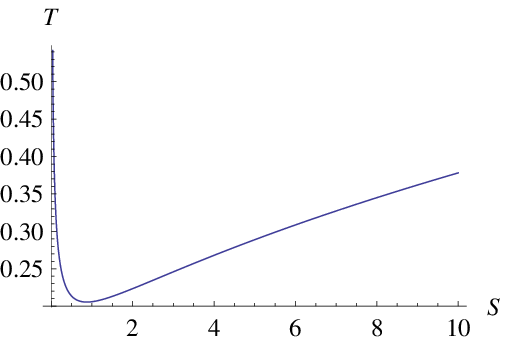}}
\subfigure[]{\label{1b}
\includegraphics[width=8cm,height=6cm]{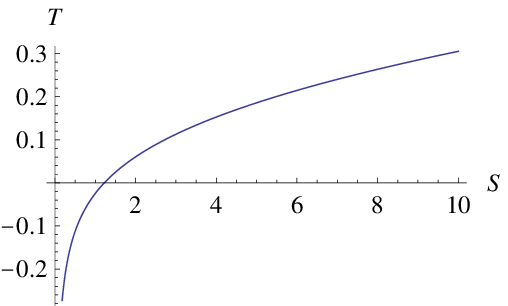}}}
 \caption{(a)Hawking temperature $T$ vs. $S$ for $b=1.5,\Phi=1,R_0=-12$ (b)Hawking temperature $T$ vs. $S$ for $b=1.5,\Phi=2,R_0=-12$;}
\label{fg1}
\end{figure}

The Hawking temperature for both the case $0<\Phi<b$ and $\Phi>b$ is depicted in Fig.~\ref{1a} and ~\ref{1b} respectively. As shown in Fig.~\ref{1a}, there exists minimum temperature when $0<\Phi<b$. Substituting Eq.(\ref{12}) into Eq.(\ref{9}), the minimum temperature can be obtained as
\begin{equation}
T_{min}=\frac{\sqrt{-R_0(b^2-\Phi^2)}}{4b\pi} .\label{14}
\end{equation}
However, the Hawking temperature increases monotonically when $\Phi>b$, as can be witnessed in Fig.~\ref{1b}.

%%%%%%%%%%%%%%%%%%%%%%%%%%%%%%%%%%%%%%%%%%%%%%%%%%%%%%%%%%%%%%%%%%%%%%%%%%%%%
%%%%%%%%%%%%%%%%%%%%%%%%%%%%%%%%%%%%%%%%%%%%%%%%%%%%%%%%%%%%%%%%%%%%%%%%%%%%%%%%

 When the electric potential of $f(R)$ AdS black hole is fixed, the specific heat can be derived as
\begin{equation}
C_\Phi=T(\frac{\partial S}{\partial
T})_\Phi=\frac{2S(-4b^2\pi+bR_0 S+4\pi \Phi^2)}{4b^2\pi+bR_0 S-4\pi \Phi^2}.\label{15}
\end{equation}
Note that the denominator of Eq.(\ref{15}) is exactly the same as the numerator of Eq.(\ref{10}), implying that the divergence of $C_\Phi$ corresponds to the minimum Hawking temperature.

When $b=1,R_0=-12/l^2=4\Lambda$, Eq.(\ref{15}) reduces to
\begin{equation}
C_\Phi=\frac{2S[\pi(1-\Phi^2)-\Lambda S]}{-\Lambda S-\pi(1-\Phi^2)},\label{98}
\end{equation}
reproducing the result of RN-AdS black holes \cite{Banerjee1}.

Substituting Eq.(\ref{7}) into Eq.(\ref{15}), one can express the specific heat into the function of $r_+$ as
\begin{equation}
C_\Phi=\frac{2b r_+^2(-4b^2\pi+bR_0 \pi r_+^2+4\pi \Phi^2)}{4b^2+b^2R_0  r_+^2-4 \Phi^2}.\label{16}
\end{equation}

One can also derive the analog of volume expansion coefficient and isothermal compressibility coefficient as
\begin{eqnarray}
\alpha&=&\frac{1}{Q}(\frac{\partial Q}{\partial
T})_\Phi=\frac{-16b^2\pi r_+}{4b^2+b^2R_0r_+^2-4\Phi^2},\label{17}
\\
\kappa_T&=&\frac{1}{Q}(\frac{\partial Q}{\partial
\Phi})_T=\frac{4b^2+b^2R_0r_+^2-12\Phi^2}{\Phi(4b^2+b^2R_0r_+^2-4\Phi^2)}.\label{18}
\end{eqnarray}
Comparing Eqs.(\ref{17}) and (\ref{18}) with Eq.(\ref{16}), one can find that both $\alpha$ and $\kappa_T$ share the same factor $4b^2+b^2R_0r_+^2-4\Phi^2$ in their denominators as the specific heat.

It is not difficult to find the condition corresponding to the divergence of $C_\Phi$, $\alpha$ and $\kappa_T$ as
\begin{equation}
4b^2+b^2R_0  r_+^2-4 \Phi^2=0,\label{19}
\end{equation}
which can be analytically solved as
\begin{equation}
r_+=\frac{2}{b}\sqrt{\frac{\Phi^2-b^2}{R_0}}.\label{20}
\end{equation}
Considering the restrictions $b>0, R_0<0$, the above root make sense physically only when $0<\Phi<b$.

\begin{figure*}
\centerline{\subfigure[]{\label{2a}
\includegraphics[width=8cm,height=6cm]{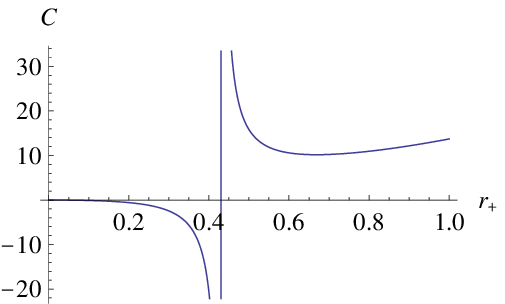}}
\subfigure[]{\label{2b}
\includegraphics[width=8cm,height=6cm]{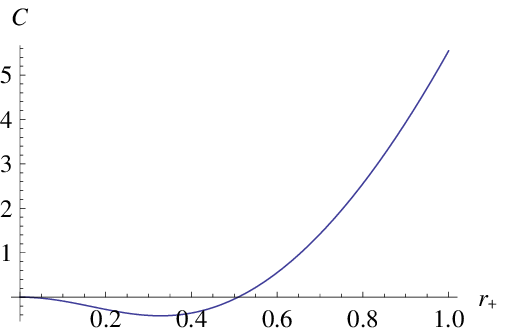}}}
\centerline{\subfigure[]{\label{2c}
\includegraphics[width=8cm,height=6cm]{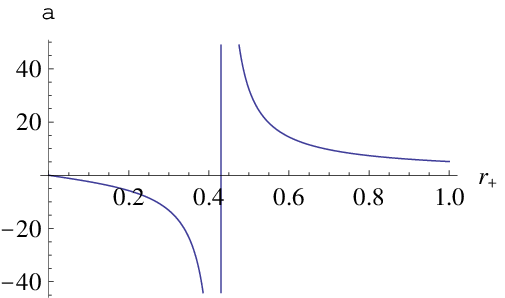}}
\subfigure[]{\label{2d}
\includegraphics[width=8cm,height=6cm]{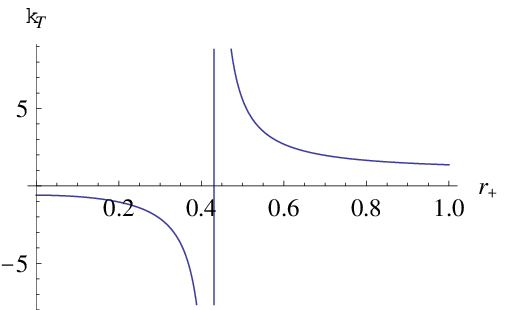}}}
 \caption{(a) $C_\Phi$ vs. $r_+$ for $R_0=-12, b=1.5, \Phi=1$ (b)  $C_\Phi$ vs. $r_+$ for $R_0=-12, b=1.5, \Phi=2$  (c)  $\alpha$  vs. $r_+$ for $R_0=-12, b=1.5, \Phi=1$  (d)  $\kappa_T$ vs. $r_+$ for $R_0=-12, b=1.5, \Phi=1$ } \label{fg2}
\end{figure*}

Fig.~\ref{2a} shows the case of $0<\Phi<b$ while Fig.~\ref{2b} shows the case of $\Phi>b$. One can see clearly that the specific heat $C_\Phi$ encounters a divergence when $0<\Phi<b$ while there is no divergence for the case $\Phi>b$. This finding also differs from our former research in the canonical ensemble~\cite{xiong6}, where there may be two, one or no divergence points for the specific heat $C_Q$.

Fig.~\ref{2c} and ~\ref{2d} show that $\alpha$ and $\kappa_T$ diverge at the same place where $C_\Phi$ does, in accordance with the above deductions.

%%%%%%%%%%%%%%%%%%%%%%%%%%%%%%%%%%%%%%%%%%%%%%%%%%%%%%%%%%%%%%%%%%%%%%%%%%%%%
%%%%%%%%%%%%%%%%%%%%%%%%%%%%%%%%%%%%%%%%%%%%%%%%%%%%%%%%%%%%%%%%%%%%%%%%%%%%%%%%

\section{Thermodynamic geometry of $f(R)$ AdS black hole}
\label{sec:4}
To examine the phase structure newly found in Sec.~\ref {sec:3}, we would like to appeal to thermodynamic geometry tools, such as Weinhold geometry~\cite{Weinhold} and Ruppeiner geometry~\cite{Ruppeiner}, which has found various applications in probing the phase structures of black holes~\cite{Ruppeiner2}-\cite{jiexiong3}.

   Weinhold's metric~\cite{Weinhold} was proprosed as
\begin{equation}
g_{ij}^{W}=\frac{\partial ^2 U(x^i)}{\partial x^i
\partial x^j}.\label{21}
\end{equation}

Utilizing Eqs.(\ref{2}), (\ref{4}) and (\ref{7}), one can express the mass of the black hole as the function of $S$ and $Q$ as
\begin{equation}
M=\frac{12b^2\pi^2Q^2+12b\pi S-R_0S^2}{24\pi^{3/2}\sqrt{bS}}.\label{22}
\end{equation}

\begin{figure}
\centerline{\subfigure[]{\label{3a}
\includegraphics[width=8cm,height=6cm]{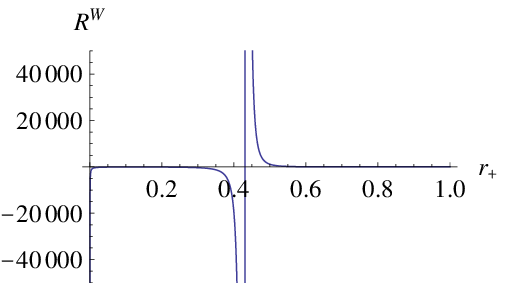}}
\subfigure[]{\label{3b}
\includegraphics[width=8cm,height=6cm]{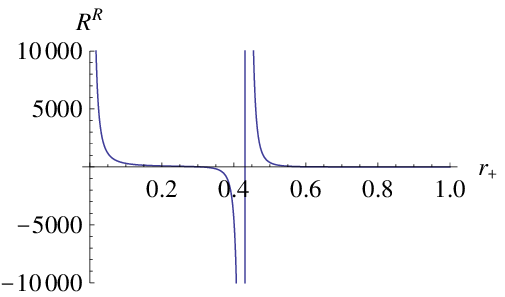}}}
 \caption{(a)Weinhold scalar curvature $R$ vs. $r_+$ for $b=1.5,\Phi=1,R_0=-12$ (b)Ruppeiner scalar curvature $R$ vs. $r_+$ for $b=1.5,\Phi=1,R_0=-12$;}
\label{fg3}
\end{figure}

Then the components of Weinhold's metric can be calculated as
\begin{eqnarray}
g_{SS}^{W}&=&\frac{12b^2\pi^2Q^2-4b\pi S-R_0S^2}{32\pi^{3/2}S^{5/2}\sqrt{b}},\label{23}
\\
g_{QQ}^{W}&=&\frac{b^{3/2}\sqrt{\pi}}{\sqrt{S}},\label{24}
\\
g_{SQ}^{W}&=&g_{QS}^{W}=\frac{-b^{3/2}\sqrt{\pi}Q}{2S^{3/2}}.\label{25}
\end{eqnarray}%
And Weinhold scalar curvature can be obtained via programming as
\begin{equation}
R^{W}=\frac{8\pi^{3/2}S^{3/2}\sqrt{b}\left[96b^3\pi^3Q^2-4b^2\pi^2(8+Q^2R_0)S-20b\pi R_0S^2-3R_0^2S^3\right]}{(4b^2\pi^2Q^2-4b\pi S-R_0S^2)^3},\label{26}
\end{equation}%
which can be reexpressed into the function of $\Phi$ as
\begin{equation}
R^{W}=\frac{8b^5(4+r_+^2R_0)(8+3r_+^2R_0)+32b^3(-24+r_+^2R_0)\Phi^2}{r_+(4b^2+b^2r_+^2R_0-4\Phi^2)^3}.\label{27}
\end{equation}%
Comparing Eq.(\ref{27}) with Eq.(\ref{16}), one may find that Weinhold scalar curvature shares the same factor $4b^2+b^2r_+^2R_0-4\Phi^2$ in the denominator as
the specific heat does, implying it would diverge exactly where the specific heat diverges. This is also shown intuitively in Fig.~\ref{3a}.

%%%%%%%%%%%%%%%%%%%%%%%%%%%%%%%%%%%%%%%%%%%%%%%%%%%%%%%%%%%%%%%%%%%%%%%%%%%%%
%%%%%%%%%%%%%%%%%%%%%%%%%%%%%%%%%%%%%%%%%%%%%%%%%%%%%%%%%%%%%%%%%%%%%%%%%%%%%%%%

Since the Ruppeiner's metric is conformally connected to the Weinhold's metric through the map \cite{Janyszek}
\begin{equation}
dS^2_R=\frac{dS^2_W}{T}.\label{28}
\end{equation}
it is convenient to derive the components of Ruppeiner's metric from those of Weinhold's metric. They can be calculated as
\begin{eqnarray}
g_{MM}^{R}&=&\frac{-12b^2\pi^2Q^2+4b\pi S+R_0S^2}{2S(4b^2\pi^2Q^2-4b\pi S+R_0S^2)},\label{29}
\\
g_{QQ}^{R}&=&\frac{-16b^2\pi^2S}{4b^2\pi^2Q^2-4b\pi S+R_0S^2},\label{30}
\\
g_{MQ}^{R}&=&g_{QM}^{R}=\frac{8b^2\pi^2Q}{4b^2\pi^2Q^2-4b\pi S+R_0S^2}.\label{31}
\end{eqnarray}%
When $b=1,R_0=-12/l^2=4\Lambda$, Eqs.(\ref{29})-(\ref{31}) reduces to
\begin{eqnarray}
g_{MM}^{R}&=&\frac{-3\pi^2Q^2+\pi S+\Lambda S^2}{2S(\pi^2Q^2-\pi S+\Lambda S^2)},\label{97}
\\
g_{QQ}^{R}&=&\frac{-4\pi^2S}{\pi^2Q^2-\pi S+\Lambda S^2},\label{96}
\\
g_{MQ}^{R}&=&g_{QM}^{R}=\frac{2\pi^2Q}{\pi^2Q^2-\pi S+\Lambda S^2},\label{95}
\end{eqnarray}%
which are equivalent to that in literature \cite{Banerjee2} of RN-AdS black holes .

Ruppeiner scalar curvature can be derived via programming as
\begin{equation}
R^{R}=\frac{A(S,Q)}{(4b^2\pi^2Q^2-4b\pi S-R_0S^2)^3(4b^2\pi^2Q^2-4b\pi S+R_0S^2)},\label{32}
\end{equation}%
where
\begin{eqnarray}
A(S,Q)&=&-1280b^7\pi^7Q^6+64b^6\pi^6Q^4(8-7Q^2R_0)S+128b^5\pi^5Q^2(6+Q^2R_0)S^2+16b^4\pi^4Q^2R_0(20-3Q^2R_0)S^3 \nonumber
\\
&\;&-336b^3\pi^3Q^2R_0^2S^4+4b^2\pi^2R_0^2(4-9Q^2R_0)S^5+16b\pi R_0^3S^6+3R_0^4S^7.\label{33}
\end{eqnarray}%
It can be reexpressed into the function of $\Phi$ as
\begin{equation}
R^{R}=\frac{B(r_+,\Phi)}{\pi r_+^2(4b^2+b^2r_+^2R_0-4\Phi^2)^3(-4b^2+b^2r_+^2R_0+4\Phi^2)},\label{34}
\end{equation}%
where
\begin{eqnarray}
B(r_+,\Phi)&=&-b^7r_+^4R_0^2(4+r_+^2R_0)(4+3r_+^2R_0)+4b^5\Phi^2[-192+r_+^2R_0(-80+84r_+^2R_0+9r_+^4R_0^2)] \nonumber
\\
&\;&+16b^3\Phi^4[-32+r_+^2R_0(-8+3r_+^2R_0)]+64b\Phi^6(20+7r_+^2R_0).\label{35}
\end{eqnarray}%
Comparing Eq.(\ref{34}) with Eq.(\ref{16}), one may find that Ruppeiner scalar curvature shares the same factor $4b^2+b^2r_+^2R_0-4\Phi^2$ in the denominator as
the specific heat does, implying it would diverge where the specific heat diverges. It can also be witnessed from Fig.~\ref{3b}. Our study of $f(R)$ AdS black holes proves again the Ruppeiner metric provides a excellent tool to probe the phase structures of black holes.

\section{Conclusions}
\label{sec:5}
  In this paper, we investigate the phase transition of four-dimensional charged AdS black hole solution in the $R+f(R)$ gravity with constant curvature in the grand canonical ensemble. It is shown that the thermodynamics in the grand canonical ensemble is quite different from that in canonical ensemble \cite{xiong6}. There exists minimum temperature when $0<\Phi<b$ while the Hawking temperature increases monotonically when $\Phi>b$. There is no critical point for $T-S$ curve, differing from the result in canonical ensemble, where we found critical point for both the $T-S$ curve and $T-r_+$ curve when the electric charge of $f(R)$ black holes is kept fixed~\cite{xiong6}.

  Moreover, we derive the explicit expression for the specific heat, the analog of volume expansion coefficient and isothermal compressibility coefficient when the electric potential of $f(R)$ AdS black hole is fixed. They share the same factor $4b^2+b^2r_+^2R_0-4\Phi^2$ in the denominator and thus share the same divergent point. The specific heat $C_\Phi$ encounters a divergence when $0<\Phi<b$ while there is no divergence for the case $\Phi>b$. This finding also differs from the result in the canonical ensemble~\cite{xiong6}, where there may be two, one or no divergence points for the specific heat $C_Q$.

To examine the phase structure of $f(R)$ AdS black hole newly found in the grand canonical ensemble, we appeal to thermodynamic geometry tools which has found various applications in probing the phase structures of black holes. We derive the analytic expressions for both the Weinhold scalar curvature and Ruppeiner scalar curvature. It is shown that they diverge exactly where the specific heat $C_\Phi$ diverges, proving again the Ruppeiner metric provides a excellent tool to probe the phase structures of black holes.

 \section*{Acknowledgements}
The authors would like to express their sincere gratitude to both the editor and the referee for their joint efforts to improve the presentation of this paper greatly. This research is supported by Guangdong Natural Science Foundation (Grant No.2015A030313789) and Department of Education of Guangdong Province of China(Grant No.2014KQNCX191). It is also supported by \textquotedblleft Thousand Hundred Ten\textquotedblright \,Project of Guangdong Province.

\end{document}